\documentclass[12pt,titlepage]{article}
\usepackage{amssymb}
\usepackage[dvips]{graphicx}
\usepackage{amsfonts}
\begin{document}
\date{}
\title{RELATIVISTIC EFFECTS OF MIXED VECTOR-SCALAR-PSEUDOSCALAR POTENTIALS FOR FERMIONS IN 1+1 DIMENSIONS}

\author{L. B. CASTRO, A. S. DE CASTRO, M. HOTT\\Departamento de F\'{i}sica e Qu\'{i}mica\\ Universidade Estadual
Paulista\\
Guaratinguet\'a, SP 12516-410, Brazil\\
benito@feg.unesp.br, castro@pesquisador.com.br, hott@feg.unesp.br}
\maketitle

\begin{abstract}
The problem of fermions in the presence of a pseudoscalar plus a
mixing of vector and scalar potentials which have equal or opposite
signs is investigated. We explore all the possible signs of the
potentials and discuss their bound-state solutions for fermions and
antifermions. The cases of mixed vector and scalar
P\"{o}schl-Teller-like and pseudoscalar kink-like potentials,
already analyzed in previous works, are obtained as particular
cases.
\end{abstract}

\section{Introduction}

The pseudospin symmetry has been used to explain a number of
phenomena in nuclear structure.\cite{bohr} In order to understand
the origin of pseudospin symmetry, we need to take into account the
motion of the nucleons in a relativistic mean field and thus
consider the Dirac equation.\cite{ginoc1} The cases in which the
mean field is composed by a vector ($V_{t}$) and a scalar ($V_{s}$)
potential, with $V_{t}=V_{s}$ ($V_{t}=-V_{s}$), are usually pointed
out as a necessary condition for occurrence of pseudospin (spin)
symmetry in nuclei.\cite{ginoc1} Furthermore, with an appropriate
choice of signs, potentials fulfilling the relations $V_{s}=\pm
V_{t}$ are able to bind either fermions or antifermions.\cite{castr}
Moreover, in the nucleus, the charge-conjugation transformation
relates the spin symmetry of the negative bound-state solutions
(antinucleons) to the pseudospin symmetry of the positive
bound-state solutions (nucleons).\cite{zhou} In addition, a link of
the pseudospin to chiral symmetry has been suggested.\cite{cohen}
The main motivation of this article is to shed some light in the
relation between spin and pseudospin symmetries by means of
charge-conjugation and chiral transformations in 1+1 dimensions. To
do so, we solve the mixed scalar-vector-pseudoscalar
P\"{o}schl-Teller potential in 1+1 dimensions, taking advantage of
the simplicity of the lowest dimensionality of the space-time as
much as was done for the harmonic oscillator.\cite%
{castro}

\section{The Dirac equation in a 1+1 dimensions}

The 1+1 dimensional time-independent Dirac equation for a fermion of
rest mass $m$ under the action of vector ($V_t$), scalar ($V_s$) and
pseudoscalar ($V_p$) potentials can be written, in terms of the
combinations $\Sigma =V_{t}+V_{s}$ and $\Delta =V_{t}-V_{s}$, as
\begin{equation}
H\Psi=E\Psi \, ,\qquad  H=c\sigma_{1} p+ \sigma_{3} mc^{2}+\frac{1+\sigma_{3} }{2}%
\Sigma +\frac{1-\sigma_{3} }{2}\Delta +\sigma_{2} V_{p},
\label{eq1a}
\end{equation}
where $\sigma_{1}, \sigma_{2}$ and $\sigma_{3}$ denote the Pauli
matrices.

The \textit{charge-conjugation operation} is accomplished by the
transformation $\Psi _{c}=\sigma_{1} \Psi ^{\ast }$ and the Dirac
equation becomes $H_{c}\Psi _{c}=-E\Psi _{c}$, with
\begin{equation}
H_{c}=c\sigma_{1}p+\sigma_{3} mc^{2}-\frac{1+\sigma_{3} }{%
2}\,\Delta -\frac{1-\sigma_{3} }{2}\,\Sigma -\sigma_{2} V_{p}\, .
\label{eq5a}
\end{equation}

The chiral operator for a Dirac spinor is the matrix $\gamma
^{5}=\sigma_{1}$. Under the \textit{discrete chiral transformation}
the spinor is transformed as $\Psi _{\chi }=\gamma ^{5}\Psi$ and the
transformed Hamiltonian $H_{\chi }=\gamma ^{5}H\gamma ^{5}$ is

\begin{equation}
H_{\chi }=c\sigma_{1} p -\sigma_{3} mc^{2}+\frac{%
1+\sigma_{3} }{2}\,\Delta +\frac{1-\sigma_{3} }{2}\,\Sigma
-\sigma_{2} V_{p}. \label{eq8}
\end{equation}%

We have two first-order equations for the upper, $\psi _{+}$, and
the lower, $\psi _{-}$, components of the spinor:
\begin{equation}
-i\hbar c\psi _{-}^{\prime }+mc^{2}\psi _{+}+\Sigma \psi
_{+}-iV_{p}\psi _{-}=E\psi _{+}  \label{eq11a}
\end{equation}%
\begin{equation}
-i\hbar c\psi _{+}^{\prime }-mc^{2}\psi _{-}+\Delta \psi
_{-}+iV_{p}\psi _{+}=E\psi _{-}\,,  \label{eq11b}
\end{equation}
\noindent where the prime denotes differentiation with respect to
$x$.

For $\Delta =0$ and $E\not=-mc^{2}$, the Dirac equation becomes
\begin{equation}
\psi_{-}= -i\bigl(\hbar c \psi^{\prime}_{+}-V_{p} \psi_{+}
\bigr)/(E+mc^{2}) \, ,\label{eq15a}
\end{equation}
\begin{equation}
-\hbar^{2}c^{2}\,\psi^{\prime \prime}_{+}\,+\left[ (E+mc^{2})\Sigma
+V^{2}_{p}+\hbar c V^{\prime
}_{p}\right]\psi_{+}=(E^{2}-m^{2}c^{4})\,\psi_{+} \, , \label{eq15b}
\end{equation}

\noindent The chiral symmetry is invoked to obtain the equations
obeyed by $\psi_{+}$ and $\psi_{-}$, for $\Sigma =0$ and
$E\not=mc^{2}$. They are obtained from the previous ones by doing
$\psi_{+} \leftrightarrow \psi_{-}$, $m \rightarrow -m$, $\Sigma
\rightarrow \Delta$ and $V_{p} \rightarrow -V_{p}$. Thus, either for
$\Delta =0$ with
$E\not=-mc^{2}$ or $\Sigma=0$ with $%
E\not=mc^{2}$ the solution of the relativistic problem is mapped
into a Sturm-Liouville problem in such a way that solutions can be
found by solving a Schr\"{o}dinger-like problem. The solutions for
$\Delta =0$ with $E=-mc^{2}$ and $\Sigma=0$ with $E=mc^{2}$%
, excluded from the Sturm-Liouville problem, are obtained directly
from the original first-order equations.

\section{The P\"{o}schl-Teller-like potential}

Let us consider

\begin{equation}
\Sigma =-\hbar c|\alpha |g_{1}\,\mathrm{sech^{2}}\alpha x,\quad
\Delta =0,\quad V_{p}=\hbar c|\alpha |g_{2}\,\mathrm{tanh}\alpha x.
\label{eq20}
\end{equation}

Due to the chiral symmetry one can restrict the discussion to the
$\Sigma $ case ($\Delta =0$). The results for the case when
$\Delta =-\hbar c|\alpha |g_{1}\,\mathrm{%
sech^{2}}\alpha x$, $\Sigma =0$, $V_{p}=\hbar c|\alpha |g_{2}\,\mathrm{tanh}%
\alpha x$ are obtained by changing the sign of $m$ and of $g_{2}$ in
the relevant expressions. The chiral symmetry can also be seen in
the isolated solutions, which are converted into each other by this
transformation.

For the potential in (\ref{eq20}) we have not found a normalizable
isolated solution with $E=-mc^{2}$ and for $E\neq -mc^{2}$, Eq.
(\ref{eq15b}) takes the form

\begin{equation}
-\frac{\hbar^{2}}{2m_{\mathtt{eff}}}\,
\psi_{+}^{\prime\prime}-U_{0}\ \mathrm{sech}^{2}\alpha x\
\psi_{+}=E_{\mathtt{eff}}\psi_{+}\, , \label{eq21}
\end{equation}
where

\begin{equation}
U_{0}=(\hbar c\alpha )^{2}\left[ g_{2}(g_{2}-1)+%
(g_{1}/\hbar c|\alpha |)(E+mc^{2})\right]/2m_{\mathtt{eff}} ,
\label{eq22}
\end{equation}%
\begin{equation}
E_{\mathtt{eff}}=(E^{2}-m_{\mathtt{eff}}^{2}c^{4})/2m_{\mathtt{eff}}\
,\qquad m_{\mathtt{eff}}=\sqrt{m^{2}+\left( \hbar \alpha g_{2}/c%
\right) ^{2}}  \label{eq23}
\end{equation}

\noindent The well-behaved solutions for Eq. (\ref{eq21}), with
$U_{0}$
necessarily real and positive, are the solutions of the Schr\"{o}%
dinger equation for the nonrelativistic symmetric modified
P\"{o}schl-Teller potential.\cite{flug} Following these previous
references and using Eqs. (\ref{eq22}) and (\ref{eq23}) we get the
quantization condition for the energy:

\begin{equation}
\hbar c|\alpha |\sqrt{1+4\left[ g_{2}(g_{2}-1)+\frac{g_{1}}{\hbar c|\alpha |}%
(E+mc^{2})\right]
}-2\sqrt{m_{\mathtt{\mathtt{eff}}}^{2}c^{4}-E^{2}}=\hbar c|\alpha
|\left( 2n+1\right) .  \label{eq25}
\end{equation}

In general there is no requirements on the signs of $g_{1}$ and
$g_{2}$, except that $U_{0}>0$. From Eq. (\ref{eq22}), the Dirac
eigenenergies corresponding to the bound-states solutions must be
within the limits

\begin{equation}
E\gtrless -mc^{2}\mp \frac{g_{2}(g_{2}-1)}{g_{1}}\hbar c|\alpha |\
,\quad \mathrm{for}\quad g_{1}\gtrless 0  \label{ranges}
\end{equation}

\noindent For $g_{1}=0$, we obtain the pseudoscalar kink-like
potential.\cite{hott2} When $g_{2}=0$ we obtain the mixed
vector-scalar P\"{o}schl-Teller-like potential with spin
symmetry.\cite{hott1} Eq. (\ref{eq25}) was solved with a symbolic
algebra program by searching eigenenergies in the ranges in
(\ref{ranges}). Fig. 1 (left) shows the behavior of the energies for
$g_{1}=-10$, for several values of $g_{2}$. For $g_{1}=10$ the
spectrum presents similar behavior if we reverse the vertical axis.
Fig. 1 (right) illustrates the behavior of the energies as a
function of $g_1$ for $g_{2}=10$.

The case $\Sigma =0$ can be obtained from the $\Delta =0$ case by
applying the charge-conjugation transformation. We recall that this
transformation
performs the changes $E\rightarrow -E$, $\Delta \leftrightarrow -\Sigma $, $%
V_{p}\rightarrow -V_{p}$, or, the changes $g_{1}\rightarrow -g_{1}$ and $%
g_{2}\rightarrow -g_{2}$. This can be seen if we solve the
eigenvalue equation for $\Sigma =0$, for $g_{1}=10$ and plot the
solutions for several values of $g_{2}$. If we compare that figure
with Fig. 1 (left) ($g_{1}=-10$) we see that the plots are identical
if we reverse both the vertical and horizontal axes, i.e, if we
reverse the sign of the energy and of $g_{2}$ concomitantly. Of
course, the identification of particle and antiparticle levels is
also reversed, as it should be, because we are applying the
charge-conjugation transformation. All of that implies into the
possibility of bounded states with both positive- and
negative-energy solutions in a system with $\Sigma=0$, that can be
relevant for nuclei.

Finally we stress that the chiral transformation enables us to
switch
between the so-called pseudospin ($\Sigma =0$ and $V_{p}=0$) and spin symmetries ($%
\Delta =0$ and $V_{p}=0$) \textit{for any potential} and, for
massless fermions, it shows that the $\Delta =0$ and $\Sigma =0$
solutions have the same spectrum. These conclusions
seem to remain true in 3+1 dimensions, because the 1+1 eigenvalue equation for $%
\Delta =0$ and $\Sigma =0$ are very similar to the corresponding 3+1
equations.

\begin{figure}[th]
\includegraphics[width=4.5cm, angle=270]{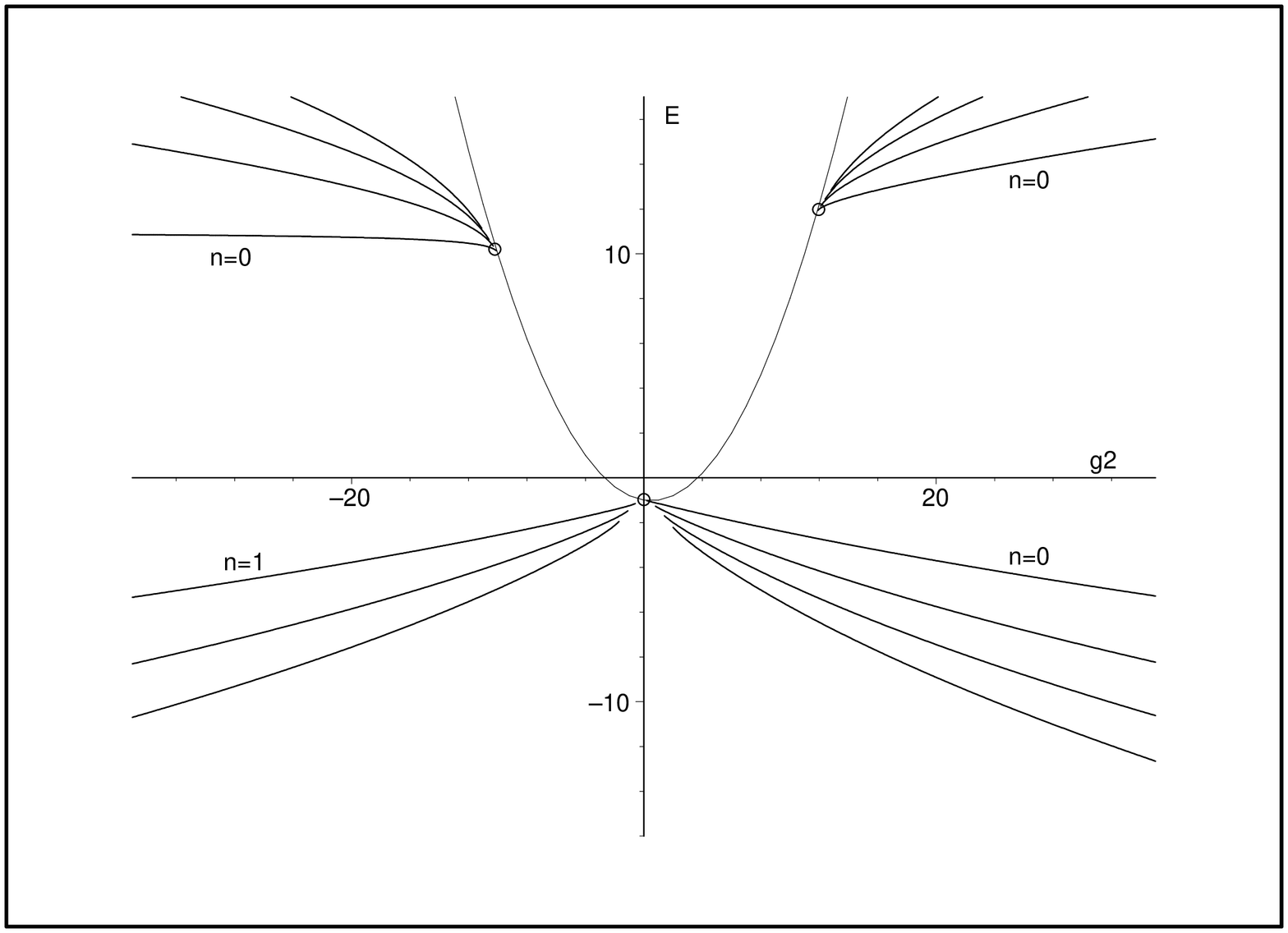}
\includegraphics[width=4.5cm, angle=270]{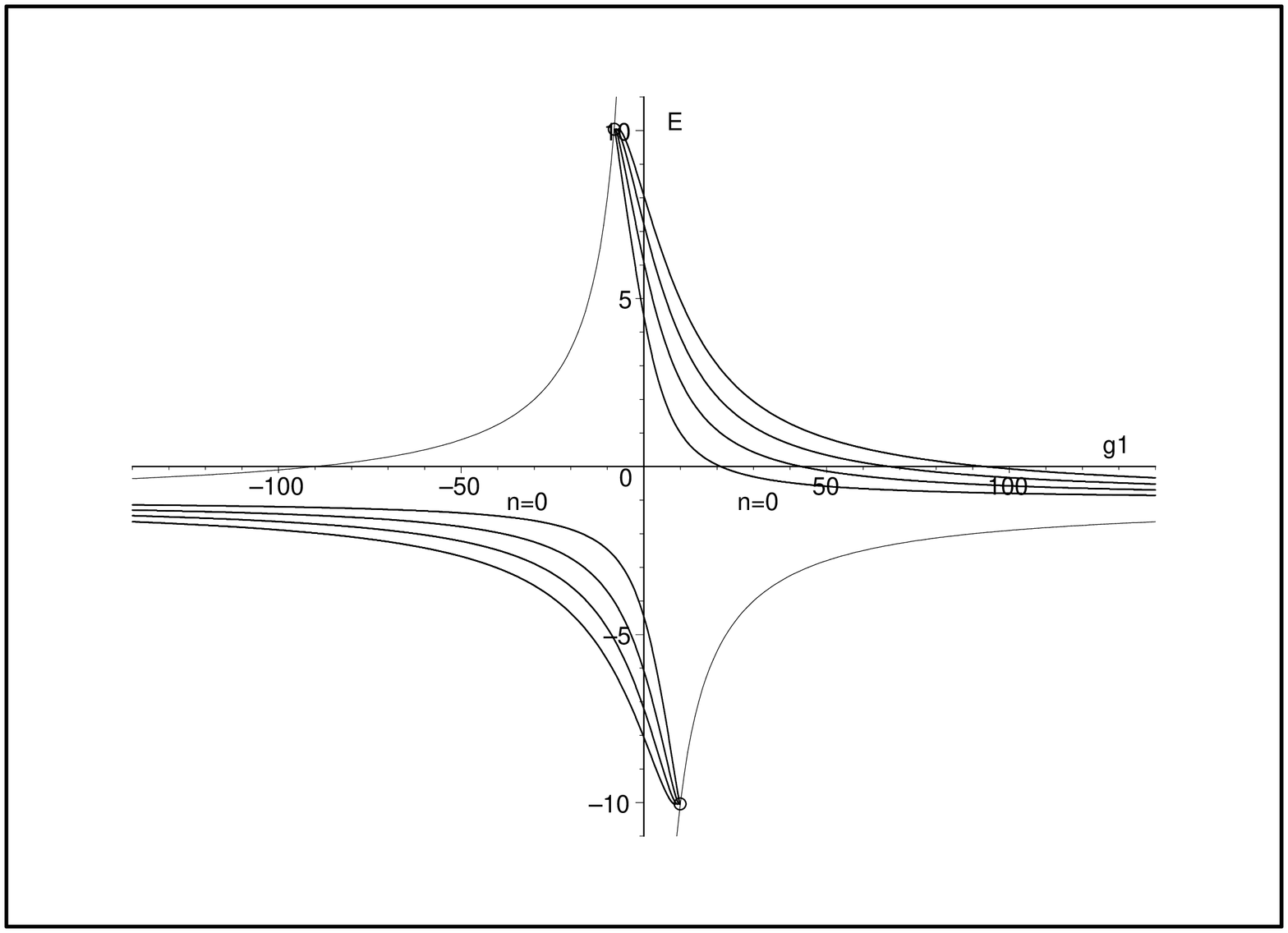}
\vspace*{-0.1cm} \caption{(left) Dirac eigenvalues for the four
lowest energy levels when $g_{1}=-10$ as a function of $g_{2}$.
(right) Dirac eigenvalues for the four lowest energy levels when
$g_{2}=10$ as a function of $g_{1}$. The thin line represents the
function $-(1+\frac{g_{2}}{g_{1}}(g_{2}-1))$ ($m=\hbar
=c=\protect\alpha =1$). }
\end{figure}
\section*{Acknowledgements}

This work is supported by CAPES, CNPq and FAPESP.


\begin{thebibliography}{0}
\bibitem{bohr} A. Bohr, I. Hamamoto, B.R. Mottelson, {\it Phys. Scr.} \textbf{26} (1982) 267. W. Nazarevicz, P.J. Twin, P. Fallon, J.D. Garrett, {\it Phys. Rev.
Lett.} \textbf{64} (1990) 1654. J.Y. Guo, X.Z. Fang, F.X. Xu, {\it
Nucl. Phys.} \textbf{A757} (2005) 411.

\bibitem{ginoc1} J.N. Ginocchio, {\it Phys. Rep.} \textbf{315} (1999)
231; \textit{ibid.} \textbf{414} (2005) 165; {\it Phys. Rev. Lett.}
\textbf{78} (1997) 436.

\bibitem{castr} A. S. de Castro, \textit{Ann. Phys. (NY)} \textbf{316} (2005) 414 .

\bibitem{zhou} S.-G. Zhou, J. Meng, P. Ring, \textit{Phys. Rev. Lett.} \textbf{91} (2003) 262501 .

\bibitem{cohen} T. D. Cohen, R. J. Furnstahl, D.
K. Griegel, \textit{Phys. Rev. Lett.} \textbf{67} (1991) 961.

\bibitem{castro} A. S. de Castro, P. Alberto, R. Lisboa, M. Malheiro,
\textit{Phys. Rev.} \textbf{C73} (2006) 054309.

\bibitem{flug} S. Fl\"{u}gge , \textit{Practical Quantum Mechanics} (Springer-Verlag,
Berlin, 1999). L.D. Landau, E.M. Lifshitz, \textit{Quantum
Mechanics} (Pergamon, N. York, 1958). M.M. Nieto, \textit{Phys.
Rev.} \textbf{A17} (1978) 1273.

\bibitem{hott2} A.S. de Castro, M. Hott, \textit{Phys. Lett.}
\textbf{A351} (2006) 379  .

\bibitem{hott1} L. B. Castro, A.S. de Castro, M. Hott, \textit{Europhys. Lett.} \textbf{77} (2007) 20009.

\end{thebibliography}
\end{document}